# An "Okay" Method for Observing Solar Eclipses

Peilong Wang[1], Jingyuan Chen[1]

[1]Department of Radiation Oncology, Mayo Clinic, Phoenix, AZ, USA

Background: Solar eclipses, as rare astronomical events, often evoke a profound sense of wonder and awe within the human spirit. However, for ordinary people, the extremely short preparation time, a few hours of notice from friends or social media, and the lack of observation equipment often hinder safe and effective eclipse viewing. Some individuals directly observe the sun with their naked eyes, risking vision damage. To enable ordinary people to safely observe eclipses in very little preparation and reduce the risk of vision damage, we present a simple and safe method that almost anyone can use under very basic conditions, known as the "Okay" observation method.

Materials and Methods: All that is needed is to find a flat wall or surface, then stand with your back to the sun, making the "Okay" 👌 gesture with your hand, allowing sunlight to pass through the "O" shape aperture formed by your thumb and index finger. This will cast the shape of the eclipse on the wall or surface for observation. This method is based on the principle of pinhole imaging. To compare the imaging quality by the "Okay" method, we also performed the observation using conventional pinhole imaging by puncturing a hole in the center of a cardboard. Both methods were tested during the partial solar eclipse observed in Phoenix, AZ on October 14, 2023.

Results: Figure 1a demonstrates that using the "Okay" observation method, the eclipse can be observed on a wall or surface with an affordable resolution. Figure 1b illustrates the eclipse observed by the cardboard pinhole method. The "Okay" observation method and cardboard pinhole imaging yield comparable results in eclipse observation, with the cardboard pinhole imaging showing less penumbra artifact.

Conclusions: The "Okay" observation method enables ordinary people to observe eclipses using only a hand gesture and a wall or surface in an extremely short preparation time, addressing the issue of safely observing eclipses without readily available tools. It also reduces the risk of vision damage that may result from incorrect direct observation with the naked eye. In the meantime, this method introduces higher penumbra artifacts due to the thickness of the fingers. Experts should seek alternative tools for higher accuracy eclipse observation.

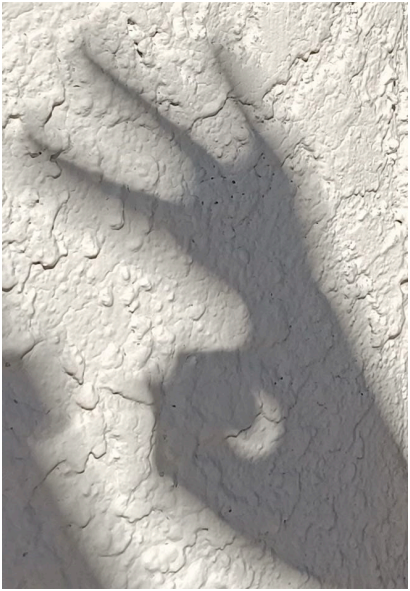 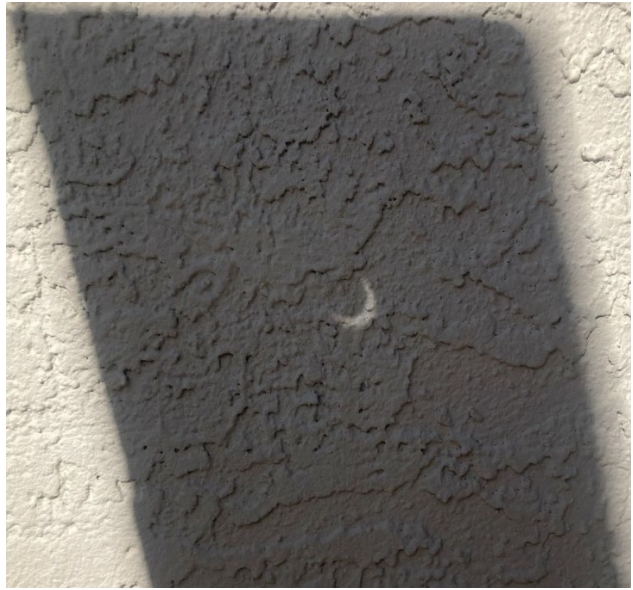

(a)                                                         (b)

Figure 1. (a) The solar eclipse observed by the "okay" method; (b) the solar eclipse observed by the cardboard pinhole method. Location: Phoenix, AZ. Date: October 14, 2023.